\documentclass[a4paper, oneside, twocolumn, notitlepage, 10pt]{extarticle_ecoc}
\usepackage{ecoc}
\usepackage{comment}
\usepackage{svg}
\begin{document}
\selectlanguage{english}    % Standard Language

%-------------------------------------------------- Title -----------------------------------------------------%

\title{Submarine Cable Deep-Ocean Observation of Mega-Thrust Earthquake and Tsunami with 44,000 100-m Spaced Sensors} %Points}%

%------------------------------------------------- Authors-----------------------------------------------------%

\vspace{-0.8cm}
\author{
    M.~Mazur\textsuperscript{(1)}, 
    N.~K.~Fontaine\textsuperscript{(1)}, 
    R.~Ryf\textsuperscript{(1)}, 
    M.~Karrenbach\textsuperscript{(2)}, %martinkarrenbach@gmail.com     
    K.~L.~McLaughlin\textsuperscript{(3)},
    B.~J.~Sperry\textsuperscript{(3)}, \\
    A.~G.~Butler\textsuperscript{(3)},
    V.~Kamalov\textsuperscript{(4)},  %vkamalov@gmail.com 
    L.~Dallachiesa\textsuperscript{(1)}, 
    E.~Burrows\textsuperscript{(1)},
    D.~Winter\textsuperscript{(1)},    
    H.~Chen\textsuperscript{(1)},  \\
    J.~Naik\textsuperscript{(5)},
    K.~Padmaraju\textsuperscript{(5)},
    A.~Mistry\textsuperscript{(5)} and    
    D.~T.~Neilson\textsuperscript{(1)}
}

\maketitle                  % Create title and author
%------------------------------------------ Description of Authors ----------------------------------------------%
\begin{strip}
\begin{author_descr}

\textsuperscript{(1)} Nokia Bell Labs, 600 Mountain Ave., Murray Hill, NJ 07974, USA.
\textcolor{blue} {\uline{mikael.mazur@nokia-bell-labs.com}} 
\\
\textsuperscript{(2)} Seismics Unusual, LLC, Brea, CA 92821, USA \\
\textsuperscript{(3)} Leidos Inc., Arlington, VA 22203, USA  \\
\textsuperscript{(4)} Valey Kamalov LLC, Gainesville, FL 32607, USA\\
\textsuperscript{(5)} Nokia Advanced Optics, 171 Madison Ave, New York City, NY 10016, USA \\
 \end{author_descr}
 \vspace{-6mm}
\end{strip}
\setstretch{1.1}
%-------------------------------------------------- Footnote -------------------------------------------------------%
\renewcommand\footnotemark{}
\renewcommand\footnoterule{}
%\let\thefootnote\relax\footnotetext{text}

%-------------------------------------------------- Abstract ---------------------------------------------------------%
\vspace{-1mm}
\begin{strip}
  \begin{ecoc_abstract}
 % NOTE: Don't use a blank line here but start abstract right away to avoid an extra line break 45 words max 
We detect the recent M8.8 mega-earthquake in Eastern Russia, on a 4400\,km long active telecom cable in the Pacific Ocean. The resolution achieved 100\,m represents the highest spatial resolution, the largest number of ocean-bottom sensors, and the first fiber-optic deep-ocean observation of a tsunami wave. 
% Corresponding author: 
% \textcolor{blue} {\uline{mikael.mazur@nokia-bell-labs.com}} 
~\textcopyright2025 The Author(s)  
 \end{ecoc_abstract}
\vspace{-3mm}
\end{strip}

%-------------------------------------------------- Introduction Section -------------------------------------------------------%
\section{Introduction} \vspace{-2mm}
\label{sec:intro}
% \begin{figure*}[b]
%     \centering
%     \textbf{Important Statement:} This is a single-column statement that appears at the bottom of the page in a two-column document. It spans both columns and is centered.
% \end{figure*}
In the 21st century there have been five mega-thrust earthquakes of magnitude 8.5 or greater\cite{GES01608}, which can have a devastating impact on people, causing damage to humans and property both from seismic waves and generated tsunamis~\cite{doocy2013human}.
% Large-scale earthquakes with a magnitude above 8 occur approximately once every year. 
% In contrast, earthquakes with a magnitude of 8.5 or larger occur on average once every decade. 
%These rare mega-earthquakes can have a devastating impact on people, causing damage to humans and property both from seismic waves and generated tsunamis~\cite{doocy2013human}. 
%These events emit strong seismic waves that shake the entire Earth, posing a unique opportunity to understand both the fault mechanism and the inner structure of the Earth on a global scale by recording the arrival times at seismic stations spread across the globe~\cite{kohler2020plan}. 
%About 70\% of earths surface is covered by oceans and 
Much of the World lacks sufficient early warning systems~\cite{allen2019earthquake} and of the 10 "grand challenges for seismology" to understand Earth's dynamics, seven require deep-ocean monitoring~\cite{lay2009seismological}. 
%Yet, $\approx73$~\% of the planet is deep ocean without any seismic sensors.
Recently, fiber optic sensing has been proposed to increase the spatial density of sensors on land as well as to increase coverage in sparsely monitored areas such as the deep ocean\cite{lindsey2021fiber,lindsey2019illuminating,williams2019distributed}. 
The use of submarine cables has been extensively investigated, but the reach of distributed acoustic sensing (DAS) has been limited to the first repeater. 
Several suggestions to extend DAS beyond the first repeater have been presented but all relies on redesign of the submarine repeaters\cite{ip2022over,ronnekleiv2024range,fan2023300}. 
\begin{figure*}[ht]
   \centering
   \vspace{-3mm}
    \includegraphics[width=1\linewidth]{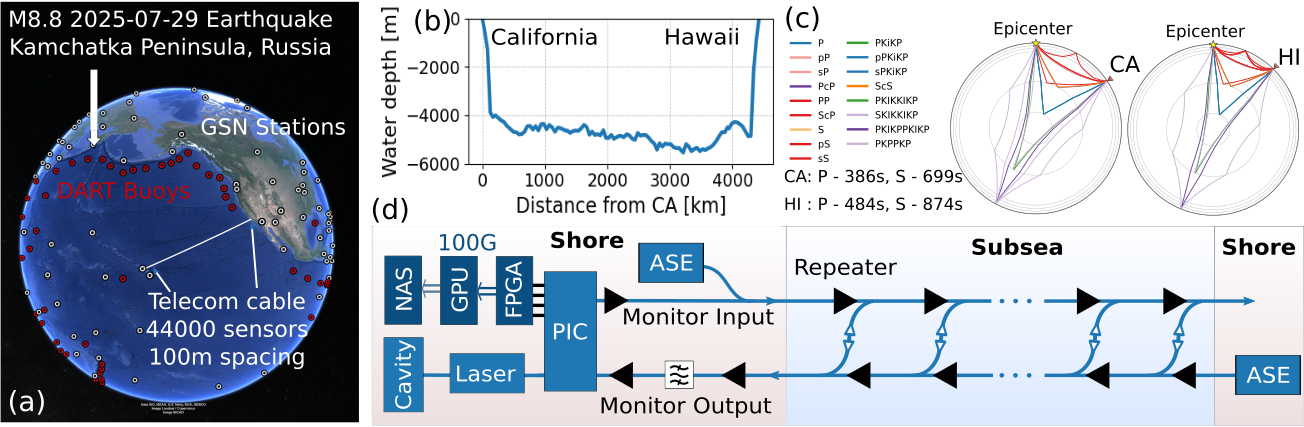}
    \caption{(a) Map showing the earthquake epicenter, the cable location, location of the DART tsunami warning buoys and the global seismic network (GSN). (b) Depth profile for the cable route. $>$ 4200\,km is at a depth exceeding 4000\,m. (c) Raytrace diagrams showing arrivals to both shore stations. All P-waves arrived at the California and Hawaii landing sites after 386s and 484s, respectively. All arrival times along the cable were within 2 minutes of each other. (d) Experimental setup to turn the submarine cable into a distributed sensor array with 100\,m spatial resolution using long-reach distributed acoustic sensing. }
    \label{fig:setup}\vspace{-1mm}
\end{figure*}
These approaches are incompatible with today's submarine cables, and therefore cannot rapidly scale since they cannot leverage the about 600 existing cables\cite{TeleGeographyMap2025}. 
Per-span interferometry which measures the phase accumulated between repeaters using the strong loopback reflectors, extends coverage to the entire length of today's transoceanic cables, but the spatial resolution is limited to 50-100\,km (e.g., the repeater spacing)\cite{marra2022optical,costa2023localization}. 
This long integration length can detect strong (M$>$6) global events, but is insufficient to resolve the fine structure of the upper layers of the earth's crust \cite{paulatto2022advances,biondi2023upper}. 
To date, earthquakes with M$\geq$8 have not been recorded using fiber optic sensing in the deep ocean. 
Furthermore, while tsunami waves have been recorded with DAS near the shore\cite{xiao2024detection,tonegawa2024high}, no tsunami has been observed beyond the first repeater, a critical feat to enable early warning of the tsunami. % should ifx

Here, we demonstrate a thousand-fold improvement in spatial resolution, relative to per-span interferometry~\cite{mazur2024global}, for fiber optic deep ocean monitoring by transforming a 4400\,km operational telecommunications submarine cable into a dense, ocean-spanning seismic array using distributed acoustic sensing (DAS). Our prototype achieves an unprecedented spatial resolution of 100\,m along the entire cable, creating a coherent network of ~44,000 sensor points with sub-repeater resolution. In a first-of-its-kind demonstration, we successfully recorded seismic waves from a rare (M$>$8.5 occur roughly once per decade) great earthquake -- a magnitude M8.8 event in Eastern Russia\cite{USGS_Kamchatka_Earthquake_2025}. 
We also demonstrate the first fiber-optic detection of a tsunami wave in the deep ocean, around 1000\,km offshore. 
This landmark observation  also validates the use of existing submarine networks for high-resolution seismic monitoring and establish a new paradigm for deep-ocean sensing, paving the way for a globally scalable network for scientific discovery, earthquake and tsunami early warning applications. %% could leave for coinclusion?
% A schematic of the experimental setup is shown in Fig.~\ref{fig:setup}.
%Mikael EDIT remaining, it's copy paste from OFC paper. 
\vspace{-3mm}
\section{Experimental Setup}\label{sec:exp}
A map of the Pacific Ocean highlighting the Earthquake epicenter and the 4400\,km long submarine cable connecting California to Hawaii is shown in Fig.~\ref{fig:setup}(a). 
In addition, white circles show the location of seismic stations part of the global seismic network (GSN) and red circles show NOAA Deep-ocean Assessment and Reporting of Tsunamis (DART)~\cite{gonzalez1998deep} buoy's deployed for tsunami early warning (ocean bottom pressure sensor). 
As can be observed, very few DART sensors are deployed and they are close to shore. Figure~\ref{fig:setup}(b) shows the cable depth profile. More than 4200\,km exceeds 4000-m of depth, out of reach for both traditional seismic networks and standard DAS, which cannot pass the repeaters. The typical span length was $\approx$45\,km and the cable had about 100 repeaters. 
Ray-trace simulations for arrival times to California and Hawaii shore stations are shown in Fig.~\ref{fig:setup}(c). All P-wave arrivals hit the cable within two minutes of each other. 
%A recording from the 2025-07-29 M8.8 mega-thrust event on the GSN station in Hawaii is shown in ~\ref{fig:setup}(c).
The experimental setup for the long-reach DAS system is shown in Fig.~\ref{fig:setup}(d). It can measure the entire cable with a spatial resolution of about 100\,m. 
The hardware consists of a photonic integrated circuit (PIC), an FPGA, a streaming-capable GPU and an NKT BASIK E15 laser locked to a vacuum reference cavity. 
The hardware is an improved version from~\cite{mazur2023advanceddistributedsubmarinecable, mazur2024global} and the software has been optimized for the weak return signal~\cite{mazur2025das}. Compatibility with live traffic over an identical cable design has been previously shown~\cite{mazur2024real}. 
The Rayleigh back-scattered signal is coupled to the shore-end via the high-loss loopback (HLLB) couplers (Fig.~\ref{fig:setup}(d)). The coupling loss was about 40\,dB and this cable does not have fiber Bragg gratings, preventing per-span interferometry~\cite{marra2022optical}. 
Dual-polarization waveforms with 250\,MHz sweep bandwidth were used and the phase change, or strain rate, was calculated using standard DAS processing very similar to ref~\cite{waagaard2021real}. 
The output was filtered and decimated to about 16\,Hz and stored on a local storage array at the cable station. This resulted in a data rate of about 3\,MB/s, which is easily manageable. 

\begin{figure*}[ht!]
   \centering
   \vspace{1.5mm}
    \includegraphics[width=1\linewidth]{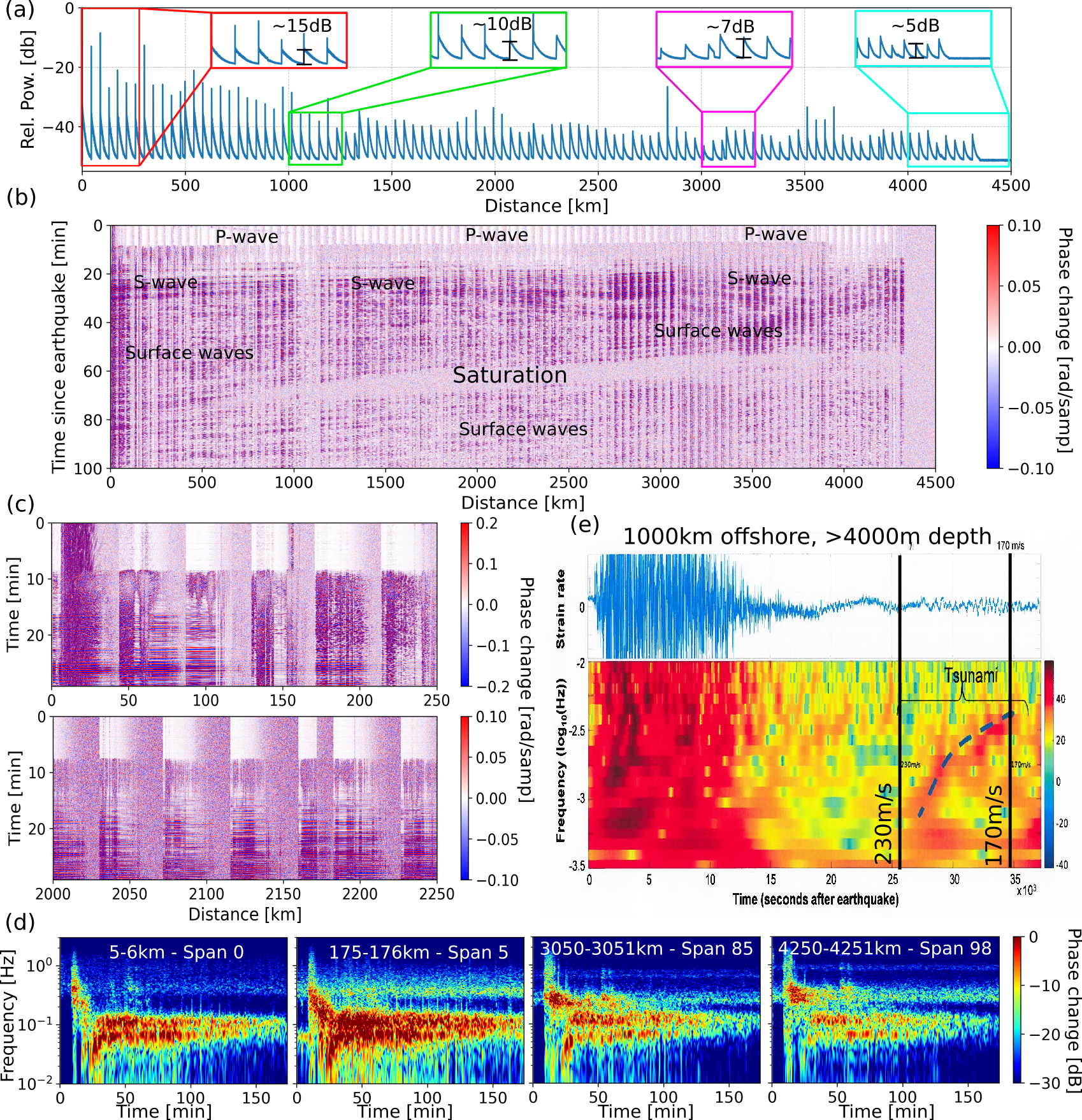}
    \caption{ (a) Loss profile with a 3\,second averaging. The SNR varied between about 15 and 5\,dB. (b) Waterfall plot showing the incoming seismic waves lighting up the sensor array. (c) Zoom-in covering the first 6 and 6 spans after 2000\,km offset. The SNR limitation towards the end of each span can easily be solved by using two systems in a bi-directional configuration. (d) Spectrograms for 4 selected distances. (e) Spectrogram showing the tsunami wave detected about 1000\,km offshore at $\approx$4400\,m depth. The tsunami wave show the strong characteristic dispersion profile with a temporal extent of about 3\,hours. It reached the cable segment around 7\,hours after the earthquake. }
    \label{fig:result}\vspace{-1mm}
\end{figure*}
\vspace{-3mm}

\section{Results}\label {Results}
\vspace{-2mm}
The measured intensity profile for a 3-second average is shown in Fig.~\ref{fig:result}(a) with insets highlighting the profile at various distances. The SNR varied between  15\,dB and 5\,dB along the fiber. The DAS signal launch power was $\approx$-3\,dBm, matching that of data channels. This was $>$10\,dB below the optimal launch power, resulting in weaker SNR. If co-existence with communications channels is necessary, the SNR penalty observed here can be recovered using two systems in a bi-directional configuration. A full waterfall trace spanning the entire cable length for a 100\,minute duration after the earthquake is shown in Fig.~\ref{fig:result}(b). Here we clearly see the P- and S-wave arrival as well as several intermediate phases. We also note that the surface waves show a clear pattern representing coherent superposition of several waves. The waveform complexity is further enhanced by the fact that there were two M$>$6 aftershocks within this time frame. We also observe some saturation around 60 minutes after the shake, which is due to $2\pi$-wrapping within the 100-meter segment originating from the very strong shaking. Spectrograms for four selected distances are show in Fig.~\ref{fig:result}(d). We observe an initial quiet environment before a clear P-wave arrival. Span 5 spectrogram shows higher signal quality compared to span 1. This is likely due to span 1 being in noisy shallow water, in contrast to around 175\,km offshore when the cable is at a water depth $>$1000-m. 
A 10 hour spectrogram integrated over 5\,km following repeater 20 is shown in Fig.~\ref{fig:result}(e). Here we observe the first detection and tracking of a tsunami passing over a communications cable beyond and between repeaters. 
%with 100 m spatial resolution utilizing OFDR. 
Coherent stacking of fifty 100\,m segments enhances the waveform in the 0.5 to 50 mHz bandwidth. The waveform dispersion just beyond the 20th repeater, about 1000\,km offshore at a depth of about 4400\,m, is shown with a continuous wavelet transform (CWT) spectrogram (amplitude vs  time and log frequency) and the band passed waveform. The onset of the tsunami was about 7 hours after the earthquake. The tsunami wave is detected on all ocean bottom spans with good coupling and can be tracked along the entire cable. 
%High-resolution DAS has the ability to separate incoming waves based on their speed. Figure~\ref{fig:result}(f) shows F-K diagrams used to analyze this (2D Fourier transforms of DAS waterfall) at 20-25\,km and 3050-3055\,km for 180 minutes following the earthquake. For span 1, we notice strong energy in the almost horizontal regime. This corresponds to $~$20\,m/s and is cause by swells. The strong dispersive waves observed on all spans are background the surface waves, traveling at speeds of about 200 to 1500-m/s. The strong earthquake waves are centered near (0,0). The near-shore observation also match similar measurements using traditional DAS~\cite{igel2024challenges}. 
\vspace{-2mm}
\section{Conclusion}\label{sec:conclusion} \vspace{-2mm}
We captured a rare, once-a-decade mega-thrust M8.8 earthquake\cite{USGS_Kamchatka_Earthquake_2025} propagating along the bottom of the Pacific Ocean using a state of the art DAS system with 4400\,km range and 100\,m spatial resolution, giving over 44,000 sensors.
Prior to this demonstration, the deep Pacific Ocean, an area larger than Earth's land mass, has been blind to any real-time high resolution seismic or strain sensing measurements. We also captured the first real-time fiber-optic detection of a tsunami in the deep ocean, more than 1000\,km offshore. 
Deploying this long-reach DAS over the entire submarine fiber network could enable new earthquake and tsunami early-warning systems and provide a paradigm-shift for large-scale scientific discoveries in the deep ocean.
% This first measurement in the deepest along the 44000 sensors 

% We observed a once-a-decade seismic event, and have demonstrated the longest high-resolution distributed acoustic sensing (DAS) measurement over a 4400\,km, repeated submarine cable, achieving 100\,m spatial resolution. We successfully captured the recent mega-thrust magnitude M8.8 earthquake in Eastern Russia\cite{USGS_Kamchatka_Earthquake_2025}, demonstrating the highest-resolution deep-ocean measurement of a large-scale earthquake to date and the first observation of an M$>$8 using long-reach fiber sensing and standard submarine cables. 
% Our results show the potential to extend DAS to the entire cable length, realizing arrays (tens of thousands) of low cost seismic and environmental sensors from a single cable system. Enabling long-reach DAS over the existing submarine network, our results paves the way for active cable protection, seismic early-warning and large-scale scientific discoveries in the deep ocean.
\vspace{2mm}
\newpage
% \footnotesize{
\onecolumn{
\printbibliography
}
%-------------------------------------------------- Bibliography Section -------------------------------------------------------%
%%%%%%%%%%%%%%%%%%%%%%%%%%%%%%%%%%%%%%%%%%%%%
%---------------------------------------------- End of Document -----------------------------------------------%
\end{document}